\documentclass{optica-article}

\journal{opticajournal} % for journals or Optica Open

\articletype{Research Article}

\usepackage[style=nature, doi=false,isbn=false,url=false,eprint=false,arxiv=false]{biblatex}
\bibliography{Main.bib}%
\AtEveryBibitem{\clearfield{month}}
\AtEveryBibitem{\clearfield{day}}
\AtEveryBibitem{\clearlist{language}}

\usepackage{lineno}
\newcommand{\LinesInFigure}{}

\newcommand{\OmegB}{\ensuremath{\Omega_\mathrm{B}}}

\usepackage{siunitx}
\usepackage{placeins}
\newcommand{\MaxStorageTimeDirect}{$114\,\mathrm{ns}$ }
\newcommand{\MaxStorageTimeCoherent}{$120\,\mathrm{ns}$ }

\begin{document}

\title{High-speed coherent photonic random-access memory in long-lasting sound waves}

\author{Andreas Geilen\authormark{1, 2, *}, Steven Becker\authormark{1, 2,*}, and Birgit Stiller\authormark{1, 2, $\dagger$}}

\address{\authormark{1} Max-Planck-Institute for the Science of Light, Staudtstr. 2, 91058 Erlangen, Germany\\
\authormark{2} Department of Physics, Friedrich-Alexander-Universität Erlangen-Nürnberg, Staudtstr. 7, 91058 Erlangen, Germany\\}
\email{\authormark{*} authors contributed equally, $\dagger$ Corresponding email: birgit.stiller@mpl.mpg.de} 

%% ==========================================================================================
%% ==========================================================================================
%% ==========================================================================================

\begin{abstract*}
In recent years, remarkable advances in photonic computing have highlighted the need for photonic memory, particularly high-speed and coherent random-access memory. Addressing the ongoing challenge of implementing photonic memories is required to fully harness the potential of photonic computing.
A photonic-phononic memory based on stimulated Brillouin scattering is a possible solution as it coherently transfers optical information into sound waves at high-speed access times. Such an optoacoustic memory has shown great potential as it fulfils key requirements for high-performance optical random-access memory due to its coherence, on-chip compatibility, frequency selectivity, and high bandwidth.  
However, the storage time has so far been limited to a few nanoseconds due to the nanosecond decay of the acoustic wave.
In this work, we experimentally enhance the intrinsic storage time of an optoacoustic memory by more than one order of magnitude and coherently retrieve optical information after a storage time of 120\,ns.
This is achieved by employing the optoacoustic memory in a highly nonlinear fiber at 4.2\,K, increasing the intrinsic phonon lifetime by a factor of six.
We demonstrate the capability of our scheme by measuring the initial and readout optical data pulse with a direct and double homodyne detection scheme.
Finally, we analyze the dynamics of the optoacoustic memory at different cryogenic temperatures in the range of 4.2\,K to 20\,K and compare the findings to continuous wave measurements.
The extended storage time is not only beneficial for photonic computing, but also for Brillouin applications that require long phonon lifetimes, such as optoacoustic filters, true-time delay networks, and synthesizers in microwave photonics.
\end{abstract*}

%% ==========================================================================================
%% ==========================================================================================
%% ==========================================================================================

\section{Introduction}
Storing information allows humans and computers to break a task down into sub tasks, solve them individually and then use those individual solutions to solve the global problem. Thus, it is key to establish deep knowledge, understand broad connections, and solve complex problems. From paper to solid-state drives, humankind has developed a variety of storage media in order to achieve this~\cite{lian_photonic_2022}. The invention of digital memory led to a powerful technology, the von Neumann computing architecture, which has been able to change the way humankind lives. However, its
separation of memory and computing units creates a performance bottleneck~\cite{lian_photonic_2022}, this is an issue for neural networks in the field of machine learning, because its sequential, digital, procedure-based programs lead to low computational efficiency and therefore to high energy consumption~\cite{shastri_photonics_2021}. 
In order to overcome the current von Neumann bottleneck, the scientific community has proposed new kinds of neuromorphic-inspired computing hardware, such as spin-based approaches~\cite{kiraly_atomic_2021}, physical systems~\cite{wright_deep_2022, dillavou_demonstration_2022}, and photonic-based architectures. 
In particular, photonic computing made huge progress over the last decade and its potential benefits 
have been demonstrated in several experiments~\cite{shen_deep_2017, zuo_all-optical_2019, lin_all-optical_2018, zhang_optical_2021, feldmann_all-optical_2019, ashtiani_-chip_2022, tegin_scalable_2021, brunner_tutorial_2018, skalli_photonic_2022, chen_deep_2023, valensise_large-scale_2022}. 
The resulting paradigm shift towards photonic information processing demands the development of high performance photonic
memory~\cite{alexoudi_optical_2020, lian_photonic_2022, kari_optical_2023, shastri_photonics_2021}.

There are two types of memory technologies which must be realized all-optically: Short term volatile and long term non-volatile memories. A long term memory cell stores information permanently and can be implemented, for instance, using phase change materials (PCMs)~\cite{yamada_rapidphase_1991, rios_integrated_2015, chen_neuromorphic_2022, meng_electrical_2023}. 
In contrast, a short term memory offers high speed access and stores information temporarily in a volatile manner. Accordingly, its optical implementation would form the photonic counterpart to the often used random access memory (RAM),
employed by modern computers~\cite{alexoudi_optical_2020}.
From perspective of emerging continuous variable quantum communication and quantum computing, it is desirable to maintain coherence of the optical information throughout the memory process.
As PCMs are incoherent and too slow to act as high-performing short-term memories, several approaches to photonic short-term memories have emerged from the scientific community~\cite{alexoudi_optical_2020, kari_optical_2023, lian_photonic_2022}. 
One of those is optoacoustic memory, which uses stimulated Brillouin scattering (SBS) in order to store and retrieve optical information in and from a traveling sound wave, respectively. 
Through the phonon-photon interaction, the optoacoustic memory offers high-speed coherent information access~\cite{merklein_chip-integrated_2017, zhu_stored_2007,stiller_brillouin_2023}, frequency selective operation~\cite{stiller_cross_2019},
integrated platform~\cite{stiller_-chip_2018} and in-memory computing capabilities~\cite{becker_optoacoustic_2023}. 
Although it fulfils many key requirements for a high-performance optical random access memory~\cite{alexoudi_optical_2020}, the nanoseconds decay of the sound wave intrinsically limits the application spectrum. 
So far, only a sophisticated acoustic refreshing scheme has increased the storage time to
a few tens of nanoseconds~\cite{stiller_coherently_2020}.

Here we experimentally reach coherent retrieval of optical information after up to \MaxStorageTimeCoherent of storage time.
This is an enhancement of the intrinsic storage time of the optoacoustic memory by more than one order of magnitude in comparison to previous work \cite{merklein_chip-integrated_2017, zhu_stored_2007, stiller_-chip_2018, stiller_brillouin_2023}.
We achieve this by operating the optoacoustic memory at cryogenic temperatures and, thereby, increasing the intrinsic lifetime of the acoustic wave by a factor of six.
Finally, we analyze the dynamics of the optoacoustic memory at different cryogenic temperatures in the range of $4.2\,\mathrm{K}$ and $20 \, \mathrm{K}$ and compare the findings to continuous wave (CW) measurements.

%% ==========================================================================================
%% ==========================================================================================
%% ==========================================================================================

\section{Theoretical basics}
Optoacoustic memory is based on stimulated Brillouin scattering which coherently couples photons with acoustic
phonons and has been well-studied~\cite{kobyakov_stimulated_2010, wolff_brillouin_2021, eggleton_brillouin_2022, eggleton_brillouin_2022-1}. 
SBS is a third-order nonlinear light-matter interaction, resulting from electrostriction and the photoelastic effect. 
In the context of optoacoustic memory, we call the involved pulsed optical fields, control and data. 
The control pulses have the optical frequency $\omega_\mathrm{c}$, whereas the data pulses are up shifted in frequency by the Brillouin 
frequency $\OmegB = 2 \omega v_\mathrm{ac} n_\mathrm{eff} / \mathrm{c_0}$, giving $\omega_\mathrm{d}=\omega_\mathrm{c}+\OmegB$ with the speed of light $c_0$, the effective optical refractive index of the waveguide $n_\mathrm{eff}$ and the speed of sound $v_\mathrm{ac}$. 

The optoacoustic memory exploits SBS to link the optical and acoustic domain, in order to store information carried by the data pulses in acoustic waves traveling more than four orders of magnitude slower~\cite{zhu_stored_2007, merklein_chip-integrated_2017}, thereby achieving a temporal delay. 
The dynamics are described by the simplified coupled mode equations~\cite{zhang_quantum_2023, wolff_brillouin_2021, kobyakov_stimulated_2010, agrawal_stimulated_2013} defined in equation~\eqref{eq: OREO_CME_simplified}. 
\begin{align}
\label{eq: OREO_CME_simplified}
\begin{aligned}
      \left(\frac{n_\mathrm{eff}}{c_{0}} \partial_t + \partial_z \right) a_{\mathrm{D}} &= - \mathrm{i} \, g_\mathrm{SBS, opt} \, a_\mathrm{C} b, \\
      \left(\frac{n_\mathrm{eff}}{c_{0}}\partial_t - \partial_z\right) a_{\mathrm{C}} &= - \mathrm{i} \, g_\mathrm{SBS, opt} \, a_\mathrm{D} b^*, \\
      \left(\partial_t + v_\mathrm{ac} \partial_z + \Gamma_\mathrm{b} \right) b &= - \mathrm{i} \, g_\mathrm{SBS, ac\,} \, a_\mathrm{D} a_\mathrm{C}^*.
\end{aligned}
\end{align}
Here we define the backward traveling control field $a_\mathrm{C}$, the forward travelling field $a_\mathrm{D}$ and an acoustic wave $b$, the optoacoustic gain for the optical and acoustic field $g_\mathrm{SBS, opt}$ and $g_\mathrm{SBS, ac}$, respectively. In addition, equation~\eqref{eq: OREO_CME_simplified} uses the acoustic linewidth $\Gamma_\mathrm{b}= \pi / \left(2 \tau_\mathrm{ph}\right)$ with the phonon lifetime $\tau_\mathrm{ph}$. %old: $\Gamma_\mathrm{b}= \pi \Delta \nu$

\begin{figure}[ht]
\centering
\includegraphics[width=1\textwidth]{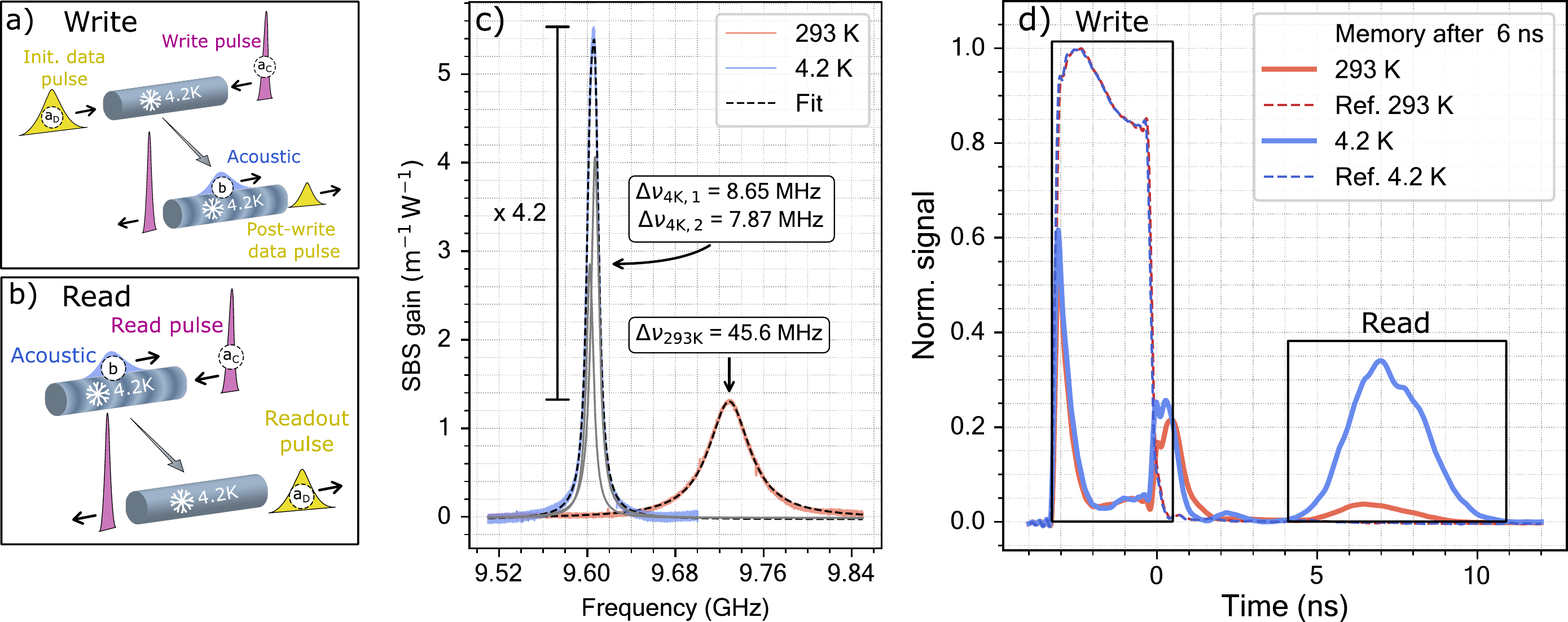}
\caption{\LinesInFigure
%suggestion
Schematic of the optoacoustic memory in an optical fiber at \SI{4.2}{\kelvin}.
\textbf{a)} Write process: Information carried by an data pulse $a_\mathrm{D}$ is stored into an acoustic wave $b$ using a strong optical write pulse $a_\mathrm{C}$.
\textbf{b)} Read process: Previously written information is transferred back into the optical domain with a strong optical read pulse $a_\mathrm{C}$.
\textbf{c)} SBS gain spectrum at $293\,\mathrm{K}$ and $4.2\,\mathrm{K}$. 
\textbf{d)} Example measurement of the optoacoustic memory, where an initial data pulse is stored for $\SI{6}{\nano\second}$. 
} 
\label{fig: Concept_and_setup}
\end{figure}

We define the storing and retrieving of the optoacoustic memory as write and read processes, respectively. 
The write process, $b(t=0)=0$, is schematically shown in Figure~\ref{fig: Concept_and_setup}\textbf{a} and employs a high-power optical write pulse $a_\mathrm{C}$  ($\omega_\mathrm{write}=\omega_\mathrm{c}$) to store the information from a counter-propagating data pulse $a_\mathrm{D}$ ($\omega_\mathrm{d}$). As soon as $a_\mathrm{C}$ and $a_\mathrm{D}$ meet in the waveguide, e.g. in the optical fiber, they induce SBS, which coherently transfers the information carried by $a_\mathrm{D}$ into an acoustic wave $b$. 
The read process, $a_\mathrm{D}(t=t_\mathrm{read})=0$, retrieves this information with a read pulse $a_\mathrm{C}$ as
depicted in Figure~\ref{fig: Concept_and_setup}\textbf{b}, analogously. 
Once $a_\mathrm{C}$ overlaps with the previously written traveling acoustic wave $b$, SBS is induced and $b$ is converted back into $a_\mathrm{D}$. 
This process recovers the amplitude and phase information carried by the initial data pulse with an intrinsic additional phaseshift of $\pi$, in accordance with~\cite{dong_photon-echo-like_2023}. 
The time delay between the write and the read pulse defines the storage time of the optoacoustic memory and is limited by the phonon lifetime. 
Hence, the memory has an intrinsic maximum storage time of several nanoseconds at room temperature, which can be increased to a few tens of nanosecond with an active refreshment scheme which amplifies the stored traveling acoustic wave~\cite{stiller_coherently_2020}.

To increase the intrinsic storage time of an optoacoustic memory by one order of magnitude, we cool down a highly nonlinear fiber (HNLF) to $T_\mathrm{HNLF} \approx4.2\,\mathrm{K}$. Details of the setup are shown in the Methods Figure~\ref{fig: full_memory_setup}.
At this temperature, the SBS gain spectrum shows a significantly higher amplitude while its linewidth is more narrow, in comparison to the room temperature spectrum (see Figure~\ref{fig: Concept_and_setup}\textbf{c}).
In fact, at temperatures up to approximately \SI{10}{\kelvin}, one observes a double peak in the spectrum (grey lines in Figure~\ref{fig: Concept_and_setup}\textbf{c}).
This is attributed to a small birefringence of the fiber that leads to two separated Brillouin frequency shifts following $\Delta \OmegB = 2 \omega v_\mathrm{ac} \Delta n_\mathrm{eff} / \mathrm{c_0}$, which only becomes visible due to the narrow linewidth. 

The measured behavior of the gain spectrum is in agreement with previously published results for a standard telecommunication single mode fiber~\cite{le_floch_study_2003}.
As for low temperatures the linewidth is inversely proportional to the phonon lifetime $\tau_\mathrm{ph}\propto\Delta\nu^{-1}$~\cite{eggleton_brillouin_2022}, 
we harness this effect to significantly increase the intrinsic storage time of the optoacoustic memory. 
This becomes clear when comparing the readout after \SI{6}{\nano\second} at room temperature (\SI{293}{\kelvin}, red line) and \SI{4.2}{\kelvin} (blue line), shown in Figure~\ref{fig: Concept_and_setup}\textbf{d}.
When normalized to the respective reference pulses (dashed lines) we observe a comparable efficiency of the writing process as the depleted pulses have the same shape and degree of depletion (see write-box in Figure~\ref{fig: Concept_and_setup}\textbf{d}).
The readout efficiency is for the \SI{4.2}{\kelvin} case significantly higher as for the room temperature case 
due to the longer phonon lifetime, resulting in a larger readout pulse 
(see read-box in Figure~\ref{fig: Concept_and_setup}\textbf{d}).

%% ==========================================================================================
%% ==========================================================================================
%% ==========================================================================================

\section{Results}
\subsection{Extension of the coherent memory time\label{sec:direct_detection}}
The extension of the phonon lifetime is observed by reading out the acoustic wave after different storage times.
With increasing delay between the write and read pulse, the readout is delayed and the amplitude decreases exponentially as the acoustic phonons decay with time constant $\tau_\mathrm{ph}$. 
For a quantitative analysis of the phonon lifetime, the complexity of the data is reduced to one parameter per memory time: the area under curve (AuC) of the readout pulse in a fixed length frame around the delay time.
This parameter is proportional to the total energy contained in the retrieved data pulse and therefore is proportional to the energy stored in the acoustic wave, which remains in the system at the readout.
We normalize the AuC to the AuC of the initial data pulse to showcase the efficiency of this process, while we observe a total writing efficiency of \SI{67}{\percent}.
Note that the degree of depletion and thus the amplitude of residual data is limited by the pump power and temporal-spectral pulse shaping and does not affect the lifetime.
\begin{figure}[hb]
\centering
\includegraphics[width=1\textwidth]{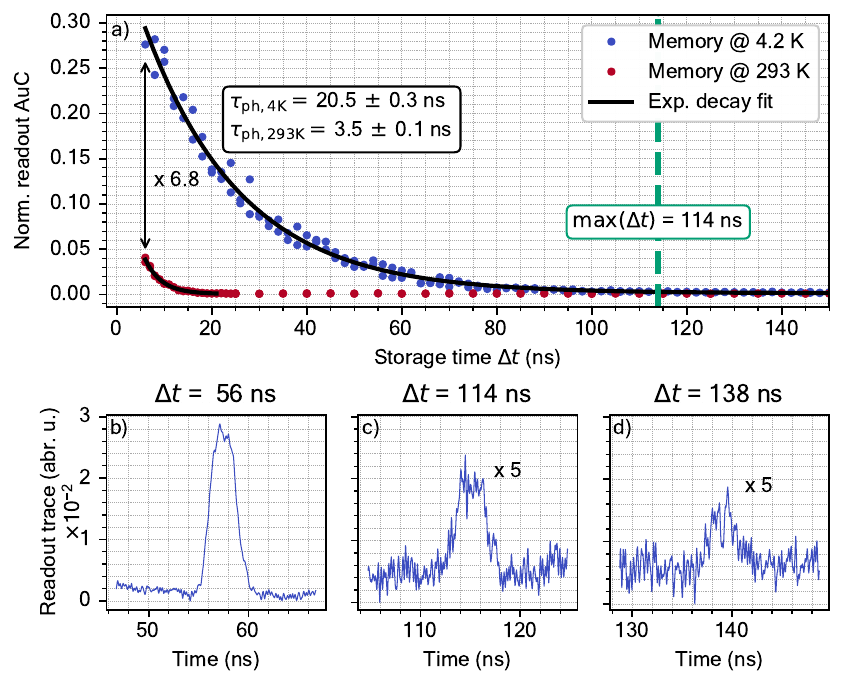}
\caption{\LinesInFigure Lifetime analysis of the cryogenic optoacoustic memory.
\textbf{a)} 
Area under curve (AuC) normalized to the AuC of the initial data pulse versus memory time $\Delta t$.
Comparison of readout at room temperature (red data) and \SI{4.2}{\kelvin} (blue data).
Increase of maximal readout efficiency by $6.8$.
Exponential fit ($AuC = A_0 \cdot \exp(-t/\tau_\mathrm{ph})) + C$) reveals 5.8 times longer intrinsic phonon lifetime at \SI{4.2}{\kelvin}.
Exemplary retrieved data pulses at \textbf{b)} \SI{56}{\nano\second}, \textbf{c)} \SI{114}{\nano\second} and \textbf{d)} \SI{138}{\nano\second}.
}
\label{fig:readout_direct_method}
\end{figure}

Like the intensity of the readout pulse, the total energy retrieved, and with it the AuC, decays exponentially.
This exponential decay of the AuC is shown in Figure~\ref{fig:readout_direct_method}~\textbf{a} for a temperature of \SI{4.2}{\kelvin} (blue curve) and room temperature (\SI{293}{\kelvin}, red curve) for comparison.
An exponential fit following $AuC = A_0 \cdot \exp(-t/\tau_\mathrm{ph}) + C$ extracts the phonon lifetime $\tau_\mathrm{ph}$, where $A_0$ is the initial AuC at $\Delta t = 0$ and $C$ is a constant linear offset.
At \SI{4.2}{\kelvin} we observe an phonon lifetime of \SI{20.5\pm0.3}{\nano\second}, which is a $5.8$ times enhancement compared to the ambient temperature value of \SI{3.5\pm 0.1}{\nano\second}.
Each point in this graph is a readout measurement.
Examples are shown in Figure~\ref{fig:readout_direct_method} \textbf{b-d}, where the initial data is averaged over $64$ traces for improved visibility.
In Figure~\ref{fig:readout_direct_method}~\textbf{b}, at a storage time of \SI{56}{\nano\second}, one can observe a clear readout at five times the timescale of the maximum passive storage time of previous work \cite{stiller_cross_2019}.
The maximal achievable memory time is \MaxStorageTimeDirect, which we define as the memory time where the signal is at least three standard deviations of the noise above the mean noise level.
The respective retrieved data pulse is shown in Figure~\ref{fig:readout_direct_method}~\textbf{c}.
At even longer memory times, we are still able to observe some readout at \SI{138}{\nano\second} as shown in Figure~\ref{fig:readout_direct_method}~\textbf{d}.
\begin{figure}[hb!]
\centering
\includegraphics[width=1\textwidth]{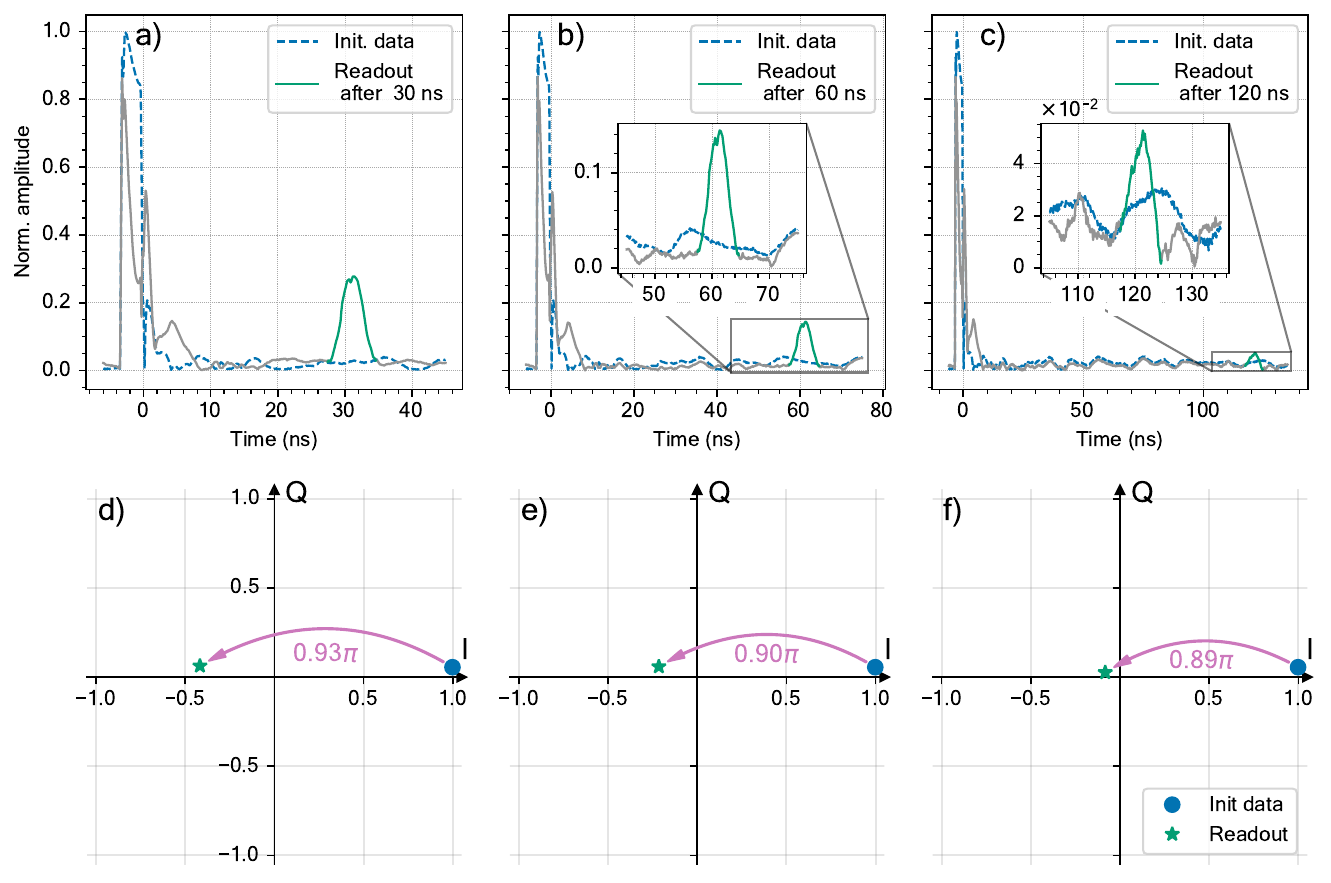}
\caption{\LinesInFigure
Fully coherent measurements of the optoacoustic memory at \SI{4.2}{\kelvin} via double homodyne detection.
Amplitude at memory time $\Delta t$ of \textbf{a)} \SI{30}{\nano\second}, \textbf{b)} \SI{60}{\nano\second} and \textbf{c)} \SI{120}{\nano\second}. 
\textbf{d) - f)} IQ representation of the initial data pulse and the retrieved data pulse at different memory times.
Phase remains constant with an additional intrinsic phase shift $\approx\pi$, demonstrating the coherence of the memory scheme at cryogenic temperatures.
\label{fig:readout_coherent}
}
\end{figure}

The Brillouin process is coherent and conserves the relative phase of the initial data pulse, analogous to previous work~\cite{merklein_brillouin-based_2018}.
We show the conservation of the phase by switching from direct detection to a double homodyne detection.
In this case, instead of the intensity of the optical signal, its amplitude and phase relative to a reference pulse is measured as the electrical voltage signal on the oscilloscope (see Method section 2). 
As a trigger for the oscilloscope, we use an additional optical pulse at frequency $\omega_\mathrm{D}$, which is not involved in the memory process. 
Figure~\ref{fig:readout_coherent} \textbf{a-c} show the oscilloscope traces of the amplitude for $64$ averages at memory times of \SI{30}{\nano\second}, \SI{60}{\nano\second} and \SI{120}{\nano\second}, respectively.
To showcase the full phase response, these traces are further shown in IQ diagrams in Figure~\ref{fig:readout_coherent}~\textbf{d-e}.
The total values of $Q$ and $I$ are determined by calculating the AuC for the channels for $I(t)$ and $Q(t)$, respectively. 
A more detailed description on how we extracted the IQ-values can be found in the supplementary material.
The data shows the constant phase of the retrieved data with the intrinsic phase shift of $\pi$ in the memory process, originating from Equation~\eqref{eq: OREO_CME_simplified} and being in agreement with~\cite{dong_photon-echo-like_2023}.
The small deviation from $\pi$ originates likely form non perfect balancing as one observes a constant positive offset in $Q$.
While the phase of the retrieved data pulse is stable, the amplitude decreases with increasing memory time analogously to the direct detection scheme.

%% ==========================================================================================
%% ==========================================================================================
%% ==========================================================================================

\subsection{Brillouin spectrum and storage time vs T}
For the investigation of the system's temperature dynamics, the experiments from the previous section are performed at temperatures from \SI{4.2}{\kelvin} (lowest possible temperature in the cryostat) to \SI{20}{\kelvin}.
In addition, seeded measurements of the integrated Brillouin gain spectrum are conducted, which gives a reference value for the phonon lifetime and determines the Brillouin frequency shift used for the temperature dependent memory measurements. A detailed description of the experiment is given in the Methods section.
\begin{figure}[ht!]
\centering
\includegraphics[width=1\textwidth]{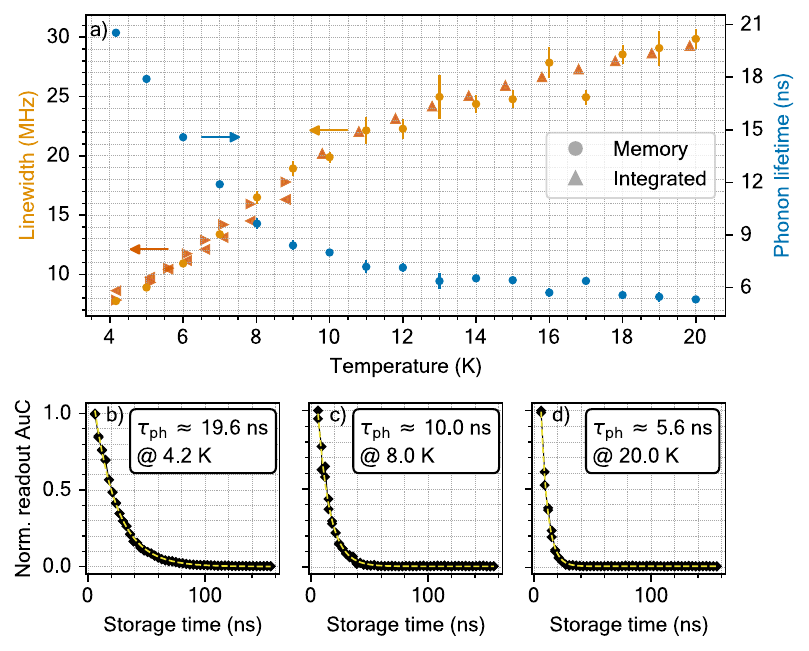}
\caption{\LinesInFigure Temperature analysis of the phonon lifetime and linewidth.
\textbf{a)}
Phonon lifetime extracted by exponential fits for temperatures from \SI{4.2}{\kelvin} to \SI{20}{\kelvin} (right y-axis).
Brillouin linewidth obtained from integrated measurements and calculated from memory via $ \Delta\nu =\ 1/\left(2\pi \tau_\mathrm{ph}\right)$ (left y-axis).
Double Lorentzian fit below \SI{10}{\kelvin} due to birefringence of the sample.
Overlap shows validity of the measurements.
Example memory sweeps at \textbf{b)} $T = \SI{4.2}{\kelvin}$, \textbf{c)} $T = \SI{8.0}{\kelvin}$, and
\textbf{d)} $T = \SI{20.0}{\kelvin}$.
}
\label{fig:T_dependence}
\end{figure}

Figure~\ref{fig:T_dependence}~\textbf{a} displays the phonon lifetime $\tau_\mathrm{ph}$ on the right y-axis versus the temperature.
Examples are shown for $T = \SI{4.2}{\kelvin}$, $T = \SI{8.0}{\kelvin}$ and $T = \SI{20.0}{\kelvin}$ in Figure~\ref{fig:T_dependence}~\textbf{b}, \textbf{c}, and \textbf{d}, respectively.
The linewidth, which is measured with the integrated setup, is plotted on the left y-axis in triangles, marked as "integrated".
The full spectra are shown in the supplementary material section 1.
It should be noted, that for temperatures below \SI{10}{\kelvin}, the linewidth becomes so narrow, that the birefringence of the sample has to be considered by fitting two Lorentzian curves, as depicted in more detail in the supplementary materials \ref{fig:T_dependence}.
For comparability, the phonon lifetimes are converted into a corresponding acoustic linewidth via $\Delta \nu =\left(2\pi\tau_\mathrm{ph}\right)^{-1}$, plotted on the left y-axis as dots as well.
One can observe a good overlap between the linewidth values obtained by the integrated and memory scheme, showing that indeed the memory can be used as a tool for direct measurement of the phonon lifetime also at cryogenic temperatures. 
In contrast to the traditional linewidth measurement method, it allows for a local analysis as only the part of the fiber where the data and control pulses overlap is probed.
In addition, this method does not suffer form the significant broadening of the spectrum at lower temperatures due to the birefringence of the fiber.
Overall, an increase of the phonon lifetime is observed for lower temperatures, in agreement with the work on single mode fiber~\cite{le_floch_study_2003}, predicting even longer phonon lifetime and, hence, memory times at even lower temperatures.

%% ==========================================================================================
%% ==========================================================================================
%% ==========================================================================================

\section{Summary and Outlook}
The present results showcase the significant extension of the memory time of the coherent optoacoustic memory to
\MaxStorageTimeCoherent by operation at cryogenic temperatures. This is an enhancement of more than one order of magnitude in comparison to previous 
works~\cite{merklein_chip-integrated_2017, zhu_stored_2007, stiller_-chip_2018, stiller_brillouin_2023}.
As this method increases the intrinsic phonon lifetime, it can be naturally combined with active refreshment schemes~\cite{stiller_coherently_2020}.

We have shown with a fully referenced double homodyne measurement, 
that the data retrieved from the optoaocoustic memory experiences a constant phase shift 
of $\approx\pi$ and maintains the relative phase to the initial data pulse, in 
agreement with previous work \cite{dong_photon-echo-like_2023}. 
As this technique measures the amplitudes of the fields instead of their intensity,
it generally benefits from double the exponential decay time. 
In our case, we where limited by the dynamic range of the balanced photodetectors. 
Our results can be beneficial to future memory applications, in particular those optical computing architectures which rely on fully-coherent information, such as~\cite{shen_deep_2017}. 

The rapid access time~\cite{stiller_brillouin_2023}, capability to be fully-integrated~\cite{merklein_chip-integrated_2017}, high-frequency selectivity~\cite{stiller_cross_2019} and
coherence of the process in combination with the enhanced storage time are not only interesting for optical computing architectures, but could leverage research on SBS and its application.
For instance, in microwave photonics~\cite{marpaung_integrated_2019} it allows one to implement narrow microwave photonic filters~\cite{marpaung_low-power_2015}, increase the delay of true-time delay networks~\cite{chin_broadband_2010,pant_photonic-chip-based_2012}, and reduce the noise level of a photonic microwave synthesizers~\cite{li_microwave_2013}. In addition, 
the recently demonstrated optoacoustic recurrent operator~\cite{becker_optoacoustic_2023} could use the higher memory time to process a larger amount of sequential data.

%% ==========================================================================================
%% ==========================================================================================
%% ==========================================================================================

\begin{backmatter}
\bmsection{Funding}
We acknowledge funding from the Max Planck Society through the Independent Max Planck Research Groups scheme and the Studienstiftung des deutschen Volkes. 

\bmsection{Acknowledgments}
T. Utikal for relentless assistance with the cryostat and vacuum components
The mechnical staff of the MPL, specifically A. Wambsganß for manufacturing individual components.
L. Blázquez Martínez and P. Wiedemann for experimental assistance.
K. Jaksch for fruitful discussions.

\bmsection{Disclosures}
The authors declare no conflicts of interest.

\bmsection{Data Availability Statement}
Data underlying the results presented in this paper are not publicly available at this time but may be obtained from the authors upon reasonable request.

\bmsection{Supplemental document}
See Supplements for supporting content. 

%% ==========================================================================================
%% ==========================================================================================
%% ==========================================================================================

\section{Methods\label{sec:method}}

\subsection{Memory measurements}
The full experimental setup is depicted schematically in Figure~\ref{fig: full_memory_setup}. A narrow linewidth laser is split up into the data and control path. The data light is up-shifted by \OmegB with a IQ-modulator used as single sideband modulator (SSM). 
Both data and control waves are pulsed into \SI{3}{\nano\second} long pulses using an arbitrary waveform generator (AWG).
The control pulses (write and read) get amplified by an high-power erbium doped fiber amplifier (HP EDFA) to an average power of approximately \SI{1.5}{\watt}. %more exact: 1.4712 +- 0.0060
The time difference between write and read pulse determine the storage time. For the storage time sweeps we vary randomly this time difference using the AWG. 
The random change in storage time ensures that slow drifting of the setup does not affect systematically the measurements. In addition, we stabilize 
the intensity modulator of the data branch with a DC-bias controller (DCB-C).
The data and control pulses are launched from opposite sides into the cryostat, where the highly nonlinear fiber (HNLF) is placed. The internal heaters of the cryostat allow to vary its temperature.
In order to ensure that the data and control pulses interact at the center part of the HNLF, where we realize the specified temperatures, there are synchronized. 
After leaving the cryostat, the data signal is then routed via a circulator into a narrow bandpass filter (NF) 
to remove reflections of the control light.
Eventually the signal is detected on a \SI{4}{\giga\hertz}-bandwidth oscilloscope (OSCI) with two different detector schemes: 
For the coherent measurements we use a double homodyne detection consisting of two \SI{2.5}{\giga\hertz} balanced photo receivers (BD) in combination with a 90 degree optical hybrid.
For direct detection (DD) and in all other measurements we use a high-speed photo diode (PD).
To determine the frequency which is used to up-shift the data frequency to the Brillouin frequency, we use the spectral data obtained by the seeded setup.
The peak of the fits determines the used Brillouin frequency.
As the system is sensitive to small deviations in the Brillouin frequency, the differing Brillouin frequency for Stokes and anti-Stokes is considered.
\begin{figure}[ht!]
\centering
\includegraphics[width=1\textwidth]{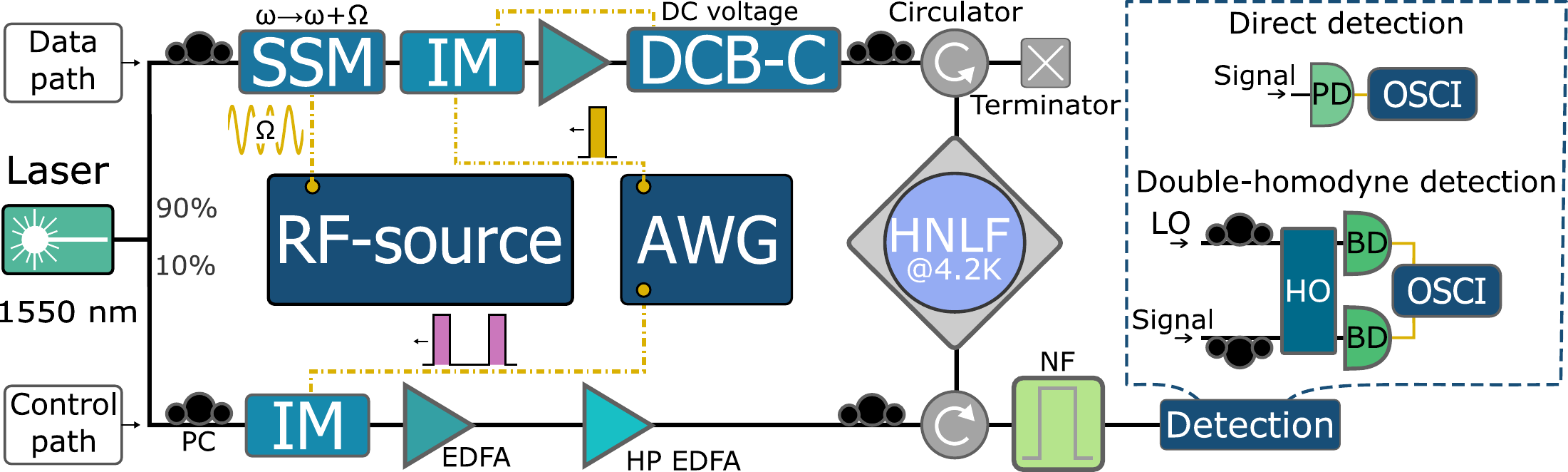}
\caption{\LinesInFigure
Experimental setup: Input laser is split into a data and control branch, where the polarization is controlled with polarization controllers (PC).
The data field is frequency-up-shifted by an IQ-modulator (SSM), driven by a high-frequency RF-source (RF-source).
Each branch is pulsed by an intensity modulator (IM). The working point of the data branch intensity modulator is additionally stabilized
with an DC-Bias controller (DCB-C). The data and contol branch uses low power erbium doped fiber amplifiers (EDFAs)
to compensate for insertion losses of the used components. 
In addition, the control branch is amplified by an high-power erbium doped fiber amplifier (HP EDFA). 
Pulses are feed into a highly nonlinear fiber (HNLF), which is placed in the cryostat. 
Data branch is filtered by a narrow bandpass filter (NF) and then launched either into direct detection (DD) or double homodyne detection scheme (DHD), measured on a $4\,\mathrm{GHz}$ oscilloscope (OSCI).
DD: High-speed photo-diode (PD). DHD: Two balanced detectors (BD) with optical 90 degree hybrid (OH) and local oscillator (LO) at the frequency $\omega_\mathrm{d}$.}
\label{fig: full_memory_setup}
\end{figure}
\subsection{Brillouin gain spectrum measurements}
In order to measure the Brillouin gain spectrum, we modify the setup from Figure~\ref{fig: full_memory_setup} 
to get the continuous wave pump-probe setup shown in Figure~\ref{fig: full_brillouin_sepectrum_setup}.
In contrast to the memory setup, we frequency down-shift the probe signal $\omega_\mathrm{probe}=\omega_\mathrm{pump}-\OmegB$, which acts as a seed for the SBS-process.
The pump branch is amplified with Erbium doped fiber amplifiers to an power of approximately \SI{71}{\milli\watt} inside the sample, considering all losses till the sample and half of the total insertion loss.
The spectrum is obtained by sweeping the frequency of the SSM in the range $\Omega_0 \le \OmegB \le \Omega_1$ and measuring the respective amplification of the seed, exemplary shown in Figure~\ref{fig: Concept_and_setup}~\textbf{c}.
First, a reference voltage curve $V_\mathrm{ref, raw}$ is measured where the pump is turned off and, thus, only the SSM-characteristic is recorded. For further data processing, we fit $V_\mathrm{ref, raw}$ with a high order-polynomial, yielding $V_\mathrm{ref}(\omega)$
After turning the pump on, the seed gets amplified according to the Brillouin gain spectrum, yielding the signal curve $V_\mathrm{sig, raw}$. In order to compensate for
small drifts in the SSM, we calculate a coefficient $c$, where we take the mean height non-amplified part of the signal curve $\overline{V}_\mathrm{sig, end}$ and the corresponding region from the reference curve $\overline{V}_\mathrm{ref, end}$ giving $c=\overline{V}_\mathrm{ref, end} / \overline{V}_\mathrm{sig, end}$.
The corrected signal curve can be noted as $V_\mathrm{sig} = c \,V_\mathrm{sig, raw}$. 
The SBS gain curve $G$ is then determined by using equation~\eqref{eq:gain_calculation}. 
\begin{equation}
    \label{eq:gain_calculation}
    G = \frac{1}{P_\mathrm{pump}L_\mathrm{cold}}\cdot \log{\left(\frac{V_\mathrm{sig}}{V_\mathrm{ref}}\right)}
\end{equation}
Where $L_\mathrm{cold}$ in equation~\eqref{eq:gain_calculation} represents the cooled part of the fiber, which is 
$L_\mathrm{cold}(T=\SI{4.2}{\kelvin})=\SI{2.0\pm 0.5}{\meter}$ and $L_\mathrm{cold}(T=\SI{293}{\kelvin})=\SI{2.4}{\meter}$. 
As the used oscilloscope does not measure the frequency of the RF-source directly, we use the RF-source's sweep output, which generates a voltage ramp, to get the frequency. Therefore, the voltage range of the ramp ($V_0$ to $V_1$) can be mapped linearly to the output frequency of the SSM ($\Omega_0$ to $\Omega_1$). As the optical signal is measured in parallel to the ramp voltage we can map $G$ to its corresponding frequency value in post-processing.
\begin{figure}[ht!]
\centering
\includegraphics[width=1\textwidth]{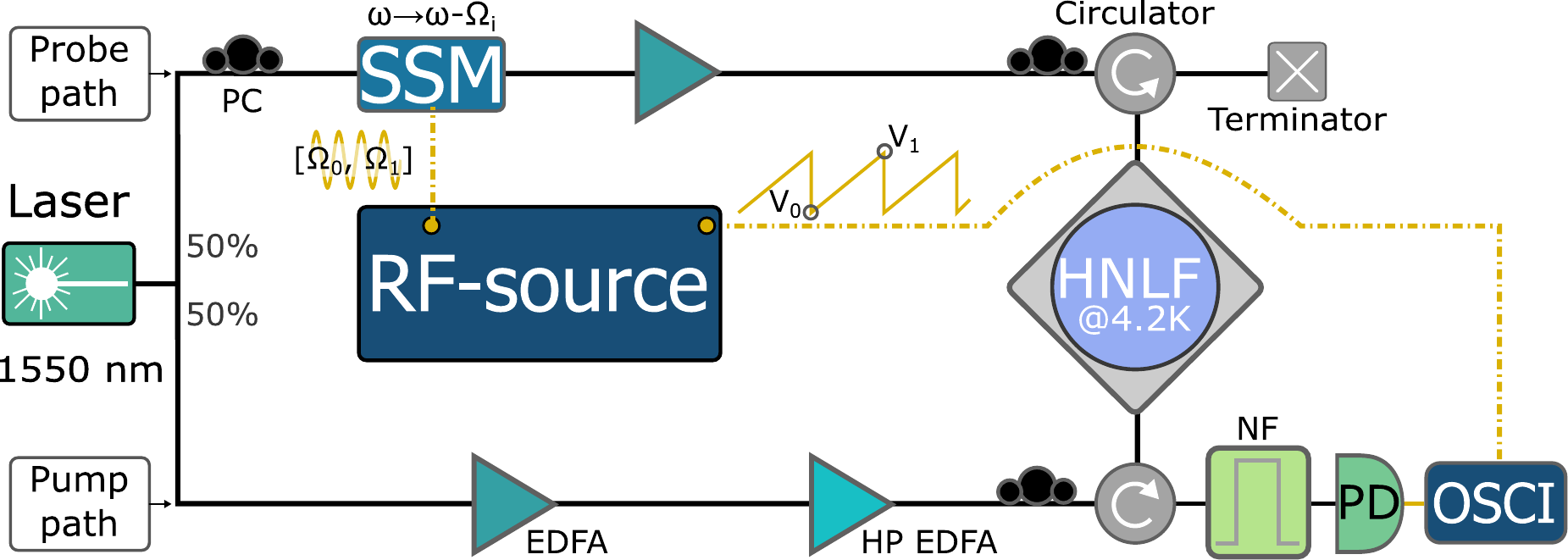}
\caption{\LinesInFigure
Experimental setup: Input laser is split into a pump and probe branch, data is frequency down-shifted by an IQ-modulator (SSB) driven by a high-frequency radio-frequency source (RF-source).
The pump is amplified by first an low-power erbium doped fiber amplifier (EDFA) and second a high-power EDFA (HP EDFA) and send into the highly nonlinear fiber (HNLF). 
Probe branch is filtered by a narrow bandpass filter (NF) and then launched either into direct detection (DD), measured on a $4\,\mathrm{GHz}$ oscilloscope (OSCI).
The frequency of the probe is swept from $\Omega_0$ to $\Omega_1$ on the RF source and the corresponding sweep voltage $V_0$ to $V_1$ is measured on the oscilloscope simultaneously.
Before the measurement the signal is optimized by changing the polarization.}
\label{fig: full_brillouin_sepectrum_setup}
\end{figure}
\end{backmatter}
\printbibliography
\end{document}

\title{Supplementary: High-speed coherent photonic random-access memory in long-lasting sound waves}

\author{Andreas Geilen\authormark{1, 2, *}, Steven Becker\authormark{1, 2,*}, and Birgit Stiller\authormark{1, 2, $\dagger$}}

\address{\authormark{1} Max-Planck-Institute for the Science of Light, Staudtstr. 2, 91058 Erlangen, Germany\\
\authormark{2} Department of Physics, Friedrich-Alexander-Universität Erlangen-Nürnberg, Staudtstr. 7, 91058 Erlangen, Germany\\}
\email{\authormark{*} authors contributed equally, $\dagger$ Corresponding email: birgit.stiller@mpl.mpg.de} 
\tableofcontents
\newpage
%% ------------------------------------------------------------------------------------------------------------------------------------------
\section{Spectra for temperature sweep}
\begin{figure}[h!]
\centering
\includegraphics[width=\textwidth]{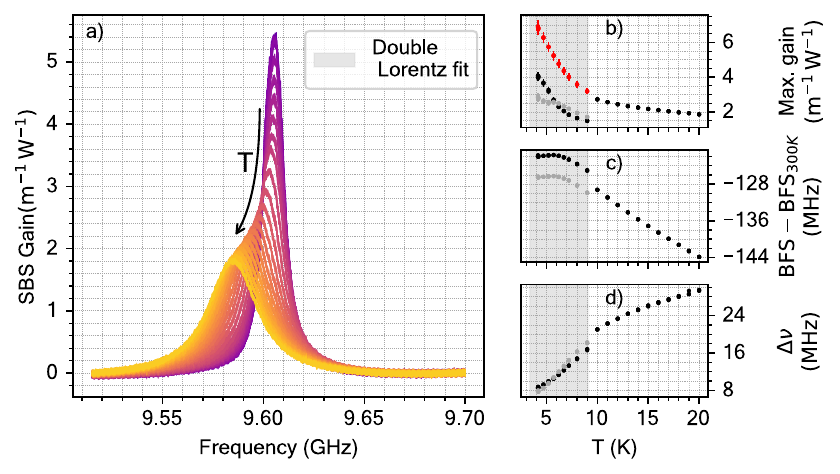}
\caption{\label{fig:spectrum_T_dependence}
\LinesInFigure
Evaluation of the integrated measurements.
\textbf{a)} Brillouin spectra with increasing temperature from \SI{4.2}{\kelvin} (purple) to \SI{20}{\kelvin} (yellow).
\textit{b) - d)} 
Fit results. 
For \SI{10}{\kelvin} and above a single Lorentzian is fitted.  
For \SI{9}{\kelvin} and below a double Lorentzian is fitted (gray area).
\textbf{b)} Gain normalized to pump power and fiber length. 
Red point are sum of the individual Lorentzian gains.
\textbf{c)} 
BFS shows the splitting of $\approx \SI{4}{\mega\hertz}$ below \SI{9}{\kelvin} most likely due to birefringence. 
This can be resolved only for low temperatures when the linewidth is small as well.
\textbf{d)} Brillouin linewidth.}
\end{figure}

We use the Brillouin gain spectrum measurement setup (see Fig. 6 of the main text) to study the Brillouin gain 
spectrum at temperatures between \SI{4.2}{\kelvin} and \SI{20}{\kelvin}. Figure~\ref{fig:spectrum_T_dependence}\textbf{a} shows the measured gain spectrum for different spectra. In order to calculate the gain, we normalize the initial spectrum with the employed pump power $\overline{P}_\mathrm{pump} = \SI{69}{\milli\watt}$ and estimated length of the cold region $L_\mathrm{cold}=\SI{2.0\pm 0.1}{\meter}$.
One can observe a movement of the spectrum and the narrowing process. In order to quantify the dynamic, we fit the spectrum with a double Lorentzian for temperatures of up to $\SI{9}{\kelvin}$ and beyond with a single Lorentzian. The reason for this is the birefringence of the sample becomes visible below $\SI{10}{\kelvin}$ as it leads to two individual peaks, separated by approximately \SI{4}{\mega\hertz}. 
Figure~\ref{fig:spectrum_T_dependence}\textbf{b} depicts the fitted gain value for the spectrum including the fitting errors and the uncertainty of the length of the cold region. The red dots represent the summed gain values of the two Lorentzian used for the double Lorentzian fit. The red dots continuously merging into the black ones supporting the double Lorentzian fit approach.
The maximum gain achieved is $g_\mathrm{max}=\SI{6.94\pm0.35}{\per\watt\per\meter}$.
Figure~\ref{fig:spectrum_T_dependence}\textbf{c} shows the Brillouin frequency shift (BFS) relative to the room temperature value $BFS_\mathrm{300K}$. 
The decreasing distance to $BFS_\mathrm{300K}$ is in agreement with~\cite{le_floch_study_2003}. 
Figure~\ref{fig:spectrum_T_dependence}\textbf{d} shows the decreasing dynamic of the Brillouin linewidth $\Delta\nu$ for lower temperatures. The lowest achieved $\Delta\nu_\mathrm{min}=\SI{7.7\pm 0.1}{\mega\hertz}$.

\newpage

%% ==========================================================================================
%% ==========================================================================================
%% ==========================================================================================

\section{Exponential decay of the acoustic amplitude}
\begin{figure}[ht!]
\centering
\includegraphics[width=1\textwidth]{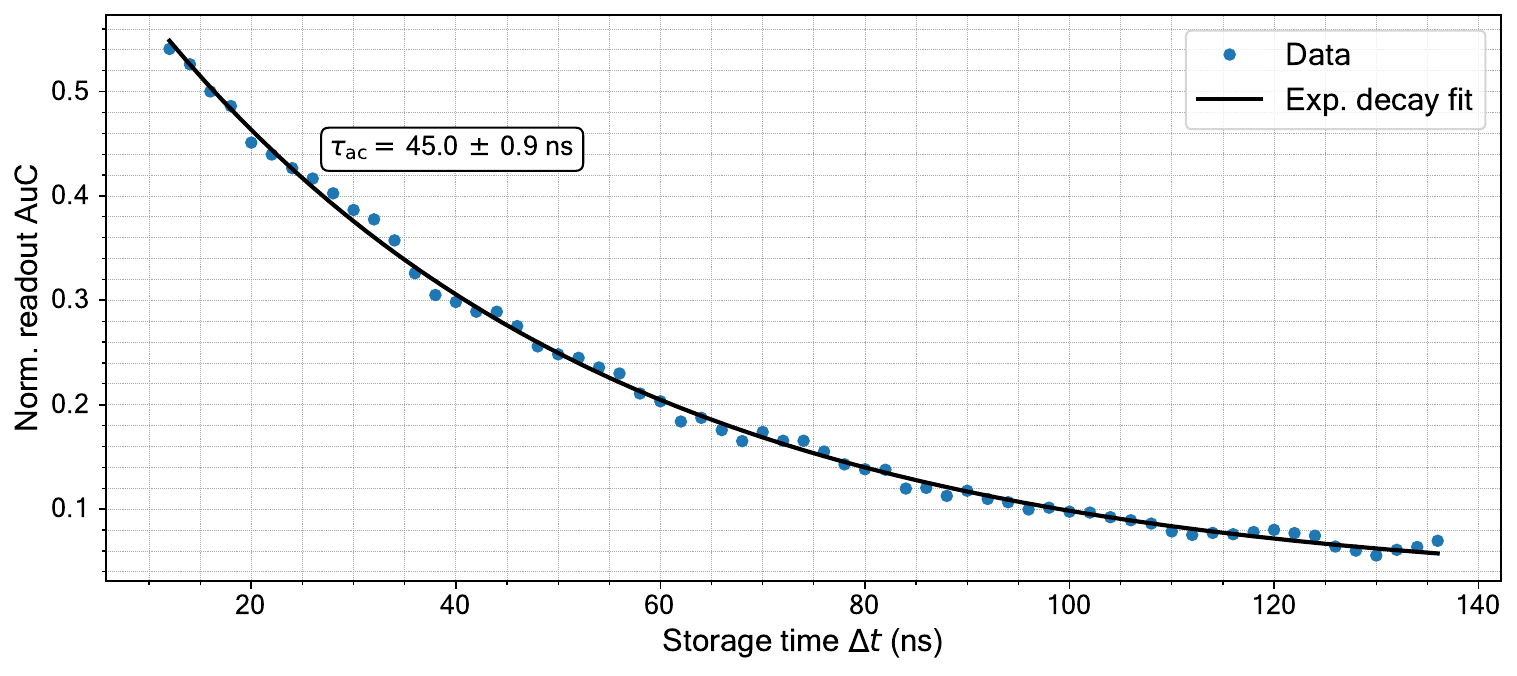}
\caption{\label{fig:exp_decay_coherent}
\LinesInFigure
Analogous to the direct detection analysis of the phonon lifetime $\tau_\mathrm{ph}$ in the main text, the lifetime of the acoustic wave $\tau_\mathrm{ac}$ can be measured with a double-homodyne detection scheme.  The exponential fit reveals $\tau_\mathrm{ac}=\SI{45.0\pm0.9}{\nano\second}$.
}
\end{figure}

In section 3.1 of the main text, we study the storage time of the optoacoustic memory measured with direct detection scheme. 
Hence, we measure the intensity of both the optical and acoustic field. As the acoustic wave $b$ decays exponentially $b(t)\propto \exp\!\left(-t/\tau_\mathrm{ac}\right)$ with the acoustic lifetime $\tau_\mathrm{ac}$, 
its intensity decays with $|b| ^2(t)\propto \exp\!\left(-2t/\tau_\mathrm{ac}\right)=
\exp\!\left(-t/\tau_\mathrm{ph}\right)$ with the phonon lifetime $\tau_\mathrm{ph} = \tau_\mathrm{ac}/2$. As consequence, an optoacoustic memory based on amplitude instead of intensity should achieve twice the storage time. In order to validate this hypothesis, we measure the area under curve (AuC) of the readout curve for different storage times $\Delta t$ with with the double-homodyne detection scheme (see Figure 5 of the main text) at \SI{4.2}{\kelvin}. Figure 3\textbf{a} to \textbf{c} of the main text show example readout traces. After collecting the AuC values, we fit them with an exponential decay $AuC(t)=A_0\exp\!\left(-t/\tau_\mathrm{ac}\right) + C$ in order to extract $\tau_\mathrm{ac}$. Figure~\ref{fig:exp_decay_coherent} shows the data and the fit. 
The amplitude of the acoustic wave decays with $\tau_\mathrm{ac}=\SI{45.0\pm0.9}{\nano\second}$ which is about two times higher as $\tau_\mathrm{ph}$ obtained in Figure 2 of the main text.   

%% ==========================================================================================
%% ==========================================================================================
%% ==========================================================================================

\section{Reference scheme for IQ data \label{IQ:reference:scheme}}
\begin{figure}[ht!]
\centering
\includegraphics[width=1\textwidth]{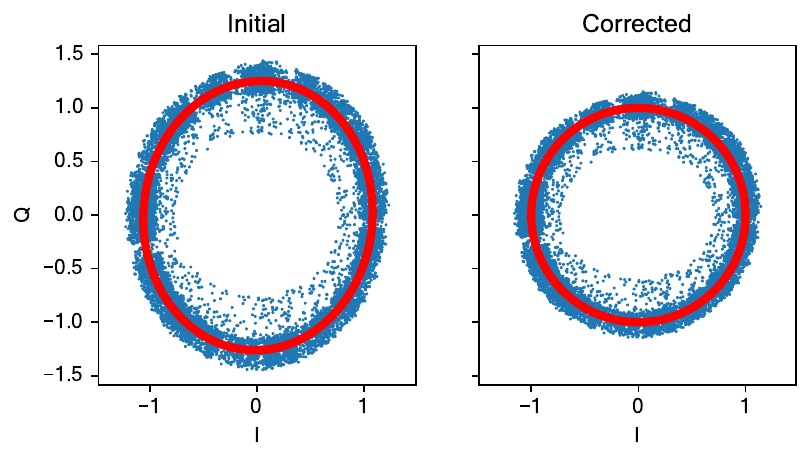}
\caption{
\label{fig:IQ_correction_result}
\LinesInFigure
Result of the $IQ$-correction. We measured the data pulse $500$ times giving the elliptical shaped data distribution (left panel). In order to correct for the elliptical distortion, we fit the distribution with an Ellipse Model (red line). After applying equation~\eqref{eq: IQ_calibration} to the initial data distribution, we get the corrected curve (right panel).}
\end{figure}

The two quadrature $I_\mathrm{cali}$ and $Q_\mathrm{cali}$ of 500 initial data pulses are measured with the oscilloscope and the double-homodyne detection scheme.
In the resulting IQ-diagram (see left panel of Figure~\ref{fig:IQ_correction_result}), 
we can see an elliptical shape instead of a circular one. 
The distortion originates from the two balanced photo detector pairs, 
which are not balanced to each other, so for both $I$ and $Q$ there is a different optical power to voltage conversion. 
In order to correct for that, we fit the data with an ellipse using the `\textsc{EllipseModel} of the \textsc{python}-package \textsc{skimage}~\cite[v0.16.2]{van_der_walt_scikit-image_2014}.
From the fit we extract the principal axes ($r_I, r_Q$), shifts of the center ($x_0, y_0)$, and the angle of rotation, which we use equation~\eqref{eq: IQ_calibration} to correct for differences in the two detectors:

\begin{align}
\label{eq: IQ_calibration}
\begin{aligned}
    I_\mathrm{corrected} &= \frac{I_\mathrm{cali}}{r_I} - x_0, \\
    Q_\mathrm{corrected} &=\frac{Q_\mathrm{cali} }{r_Q} - y_0,
\end{aligned}
\end{align}
The values for the correction are $(x_0, y_0) = (0.008, -0.005)$ and $(r_I, r_Q) = (1.069, 1.257)$. 
The result can be seen in Figure~\ref{fig:IQ_correction_result}.
We apply the same transformation for the memory data $I_\mathrm{memory}(t)$ and $Q_\mathrm{memory}(t)$.
After correcting the data, we select a rectangular time window $\tau_\mathrm{pulse}$, where the subscript $\mathrm{pulse}$ represents $\mathrm{initial}$ and $\mathrm{readout}$ for the initial data pulse and the readout pulse, respectively. With this, we define temporal mode with the corresponding $IQ$-value:

\begin{align}
\label{eq: definition_temporal_mode}
\begin{aligned}
I_\mathrm{pulse} = \int_{-\infty}^{\infty}\,\mathrm{d}t\, h_\mathrm{pulse}(t) I_\mathrm{memory}(t) \\
Q_\mathrm{pulse} = \int_{-\infty}^{\infty}\,\mathrm{d}t\, h_\mathrm{pulse}(t) Q_\mathrm{memory}(t), \\
h_\mathrm{pulse}(t) = \begin{cases}
  1,  & t \in \tau_\mathrm{pulse} \\
  0, & \mathrm{else}
\end{cases}
\end{aligned}
\end{align}

Next, we transform the $IQ$-values into polar coordinates $(a, \varphi)$ in order to extract the readout efficiency. Therefore, we use equation~\eqref{eq: IQ_2_amp_phase}.

\begin{align}
\label{eq: IQ_2_amp_phase}
\begin{aligned}
    a_\mathrm{pulse} &= \sqrt{I_\mathrm{pulse}^2 + Q_\mathrm{pulse}^2}, \\
    \varphi_\mathrm{pulse} &= \mathrm{arctan2}(I_\mathrm{pulse}, \, Q_\mathrm{pulse}) \, \mathrm{mod} \,2\pi. \\
\end{aligned}
\end{align}
We extract the read efficiency with $\tilde{a}_\mathrm{readout} = a_\mathrm{readout} / a_\mathrm{init}$. We also normalize the amplitude of the readout trace. Afterwards we transform the data back into the Cartesian coordinates, giving us the $IQ$-plot that we show in the main text. 
\clearpage
\section{References}
\printbibliography[heading=none]